\documentclass[twocolumn,pre,aps,nofootinbib]{revtex4}

\usepackage{graphicx}
\usepackage{bm}
\newcommand{\BoldVec}[1]{\mathchoice%
  {\mbox{\boldmath $\displaystyle     #1$}}%
  {\mbox{\boldmath $\textstyle        #1$}}%
  {\mbox{\boldmath $\scriptstyle      #1$}}%
  {\mbox{\boldmath $\scriptscriptstyle#1$}}%
}
\newcommand{\EQ}{\begin{equation}}
\newcommand{\EN}{\end{equation}}
\newcommand{\EQA}{\begin{eqnarray}}
\newcommand{\ENA}{\end{eqnarray}}

\newcommand{\Eq}[1]{Eq.~(\ref{#1})}

\newcommand{\Fig}[1]{Fig.~\ref{#1}}
\newcommand{\FFig}[1]{Figure~\ref{#1}}

\newcommand{\bra}[1]{\langle #1\rangle}

\newcommand{\nnn}{\hat{\mbox{\boldmath $n$}} {}}

\newcommand{\rr}{\hat{\mbox{\boldmath $r$}} {}}

\newcommand{\rrr}{\BoldVec{r} {}}
\newcommand{\xx}{\BoldVec{x}{}}

\newcommand{\uu}{\BoldVec{u} {}}

\newcommand{\ff}{\BoldVec{f} {}}

\newcommand{\FF}{\BoldVec{F} {}}

\newcommand{\nab}{\BoldVec{\nabla} {}}

\newcommand{\SSSS}{\mbox{\boldmath ${\sf S}$} {}}

\newcommand{\DD}{{\rm D} {}}
\newcommand{\dd}{{\rm d} {}}
\newcommand{\const}{{\rm const}  {}}

\def\la{\mathrel{\mathchoice {\vcenter{\offinterlineskip\halign{\hfil
$\displaystyle##$\hfil\cr<\cr\sim\cr}}}
{\vcenter{\offinterlineskip\halign{\hfil$\textstyle##$\hfil\cr<\cr\sim\cr}}}
{\vcenter{\offinterlineskip\halign{\hfil$\scriptstyle##$\hfil\cr<\cr\sim\cr}}}
{\vcenter{\offinterlineskip\halign{\hfil$\scriptscriptstyle##$\hfil\cr<\cr\sim\cr}}}}}

\def\Rey{\mbox{\rm Re}}

\def\half{{\textstyle{1\over2}}}

\def\onethird{{\textstyle{1\over3}}}

\newcommand{\yjfm}[3]{, J. Fluid Mech. {\bf #2}, #3 (#1).}

\newcommand{\ypre}[3]{, Phys.\ Rev.\ E {\bf #2}, #3 (#1).}
\newcommand{\yprl}[3]{, Phys.\ Rev.\ Lett.\ {\bf #2}, #3 (#1).}

\newcommand{\yepl}[3]{, Europhys. Lett. {\bf #2}, #3 (#1).}

\newcommand{\yjcp}[3]{, J. Comp. Phys. {\bf #2}, #3 (#1).}

\newcommand{\yapj}[3]{, Astrophys. J. {\bf #2}, #3 (#1).}

\newcommand{\ypp}[3]{, Phys. Plasmas {\bf #2}, #3 (#1).}

\newcommand{\ypf}[3]{, Phys. Fluids {\bf #2}, #3 (#1).}

\newcommand{\yjour}[4]{, #2 {\bf #3}, #4 (#1).}

\newcommand{\yproc}[4]{, (ed. #3), pp. #2. #4 (#1).}

\begin{document}

\title{Inertial range scaling in numerical turbulence with hyperviscosity}

\author{Nils Erland L.\ Haugen}
  \affiliation{Department of Physics, The Norwegian University of Science
  and Technology, H{\o}yskoleringen 5, N-7034 Trondheim, Norway}
  \email{nils.haugen@phys.ntnu.no}
\author{Axel Brandenburg}
  \affiliation{NORDITA, Blegdamsvej 17, DK-2100 Copenhagen \O, Denmark}
  \email{brandenb@nordita.dk}

\date{Received 13 February 2004; revised manuscript received 7 June 2004}

\begin{abstract}
Numerical turbulence with hyperviscosity is studied and compared with
direct simulations using ordinary viscosity and data from wind
tunnel experiments.
It is shown that the inertial range scaling is similar in all three cases.
Furthermore, the bottleneck effect is approximately equally broad (about
one order of magnitude) in these cases and only its height is
increased in the hyperviscous case--presumably as a consequence of the
steeper decent of the spectrum in the hyperviscous subrange.
The mean normalized dissipation rate is found to be in agreement with
both wind tunnel experiments and direct simulations.
The structure function exponents agree with the She-Leveque model.
Decaying turbulence with hyperviscosity still gives the usual $t^{-1.25}$
decay law for the kinetic energy, and also the bottleneck effect is
still present and about equally strong.
\end{abstract}
\maketitle

\section{Introduction}

In recent years there has been growing awareness of the detailed
structure of the kinetic energy spectrum of hydrodynamic turbulence.
In addition to the basic Kolmogorov $k^{-5/3}$ spectrum with an
exponential dissipation range there are strong indications of intermittency
corrections (possibly throughout the entire inertial range) and there is
also the so-called bottleneck effect \cite{bottleneck,LohseMuellerG},
i.e.\ a shallower spectrum near the beginning of the dissipative
subrange; see also Ref.~\cite{SJ93}.
These features can be seen both in high resolution simulations
\cite{Kan03} and in measurements of wind tunnel turbulence \cite{PKJ03}.

Over the past few years it has become evident that in
numerical turbulence the bottleneck effect is rather pronounced
\cite{PorterWoodward1998,GotohFukayama2001,Kan03}.
However, some of the simulations used hyperviscosity or other kinds of
subgrid scale modeling.
Hyperviscosity has frequently been used in turbulence studies in order
to shorten the dissipative subrange \cite{BLSB81,MFP81,McW84,PP88,BO95}.
However, hyperviscosity has also been suggested as a possible source
of an artificially enhanced bottleneck effect \cite{BSC98,BM00}.
Meanwhile, the apparent discrepancy in the strength of the bottleneck
effect between simulations and experiments
has been identified as being due to the difference in the
diagnostics:
in wind tunnel experiments one is only able to measure
one-dimensional (longitudinal or transversal) energy spectra,
while in simulations one generally considers shell integrated
three-dimensional spectra.
The two are related by a simple integral transformation
\cite{Batchelor,Hinze,MoninYaglom}.
It turns out that, while the bottleneck effect can be much weaker or
even completely absent in the one-dimensional spectrum, it is generally
much stronger in the three-dimensional spectrum \cite{DHYB03}.

In order to see the bottleneck effect in simulations, it is important
to have sufficiently large resolution of around $1024^3$ meshpoints.
This raises the question to which extent the bottleneck effect seen
in simulations with hyperviscosity is an artifact or a real feature
that becomes noticeable only above a certain resolution.
It is thus possible that the reason for an exaggerated bottleneck effect
in the hyperviscous simulation is related to the fact that
hyperviscosity increases the effective resolution beyond the threshold
above which the bottleneck effect can be seen.

In this paper we consider forced hydrodynamic turbulence using
hyperviscosity proportional to $\nabla^6$ (instead of the usual
$\nabla^2$ viscosity operator).
We find that the bottle\-neck effect is enhanced in amplitude--but
not in width, compared with direct simulations at the currently
largest resolution of $4096^3$ on the Earth Simulator \cite{Kan03}.
One of the important results of these very high resolution simulations is
that an inertial range begins to emerge that is clearly distinct
from the bottleneck effect.
Furthermore, the (negative) slope in the inertial range is steeper
than the standard Kolmogorov power law exponent of $5/3$ by about $0.1$,
so it is approximately $1.77$.

As in earlier papers \cite{DHYB03}, we consider weakly compressible
turbulence using an isothermal equation of state.
The root mean square Mach number is between $0.12$ and $0.13$; for this
type of weakly compressible simulations, we find that the energies of
solenoidal and potential components of the flow have a ratio
$E_{\rm pot}/E_{\rm sol} \approx 10^{-4}\mbox{--}10^{-2}$ for most
scales; only towards the Nyquist frequency the ratio increases to about $0.1$.
Compressibility is therefore not expected to play an important role.

\section{Basic equations}

We solve the compressible Navier-Stokes equations,
\begin{equation}
\frac{\DD\uu}{\DD t}=-{1\over\rho}\nab p+\ff+\FF_{\rm visc},
\end{equation}
where $\DD/\DD t=\partial/\partial t+\uu\cdot\nab$ is the advective
derivative, $p$ is pressure, $\rho$ is the density,
$\ff$ is an isotropic, random, nonhelical
forcing function with power in a narrow band of wavenumbers, and
\begin{equation}
 \label{22prelim}
\FF_{\rm visc}={1\over\rho}\nab\cdot
\left(2\rho\nu_n\SSSS^{(n)}\right)
\end{equation}
is the viscous force. Here,
\begin{equation}
\SSSS^{(n)}=(-\nabla^2)^{n-1}\SSSS
\end{equation}
is a higher order traceless rate of strain tensor,
\begin{equation}
{\sf S}_{ij}=\half\left(
u_{i,j}+u_{j,i}\right)-\onethird\delta_{ij}\nab\cdot\uu
\end{equation}
is the usual traceless rate of strain tensor,
and commas denote partial differentiation.
In the following we restrict ourselves to the case where
$\mu_n\equiv\rho\nu_n=\const$.
Using the product rule, we can then rewrite \Eq{22prelim} in the form
\begin{equation}
 \label{22}
\FF_{\rm visc}
=(-1)^{n-1}{\mu_n\over\rho}\left(\nabla^{2n}\uu
+\onethird\nabla^{2(n-1)}\nab\nab\cdot\uu\right).
\end{equation}
For $n=1$ we recover the normal diffusion
operator for compressible flows. In the present paper we choose $n=3$,
so equation (\ref{22}) reduces to 
\begin{equation}
\label{23}
\FF_{\rm visc}
={\mu_3\over\rho}\left(\nabla^{6}\uu
+\onethird\nabla^{4}\nab\nab\cdot\uu\right).
\end{equation}
In the incompressible case, which is usually considered, the second
term in \Eq{23} vanishes.
However, in the compressible case considered here this term is important
to ensure momentum conservation.
The local rate of kinetic energy dissipation per unit mass is
\begin{equation}
\epsilon=2\mu_3\left(\nabla^2\SSSS\right)^2,
\end{equation}
which is positive definite.

We consider an isothermal gas with constant sound speed $c_{\rm s}$,
so that the pressure is given by $p=c_{\rm s}^2\rho$ and
$\rho^{-1}\nab p=c_{\rm s}^2\nab\ln\rho$.
The density obeys the continuity equation,
\begin{equation}
\frac{\DD\ln\rho}{\DD t}=-\nab\cdot\uu.
\end{equation}
For all our simulations we have used the {\sc Pencil Code} \cite{PC}, which
is a grid based high order code (sixth order in space and third order
in time) for solving the compressible hydrodynamic equations.

\section{Results}

We have calculated a series of models with resolutions varying between
$64^3$ and $512^3$ meshpoints using a third order hyperviscosity ($n=3$).
When changing the resolution, we keep the grid Reynolds number,
here defined as
\begin{equation}
\Rey_{\rm grid}=u_{\rm rms}\left/\left(\nu_n k_{\rm Ny}^{2n-1}\right)\right.
\end{equation}
approximately constant.
Here, $k_{\rm Ny}=\pi/\delta x$ is the Nyquist wavenumber and
$\delta x$ is the mesh spacing.
Thus, when doubling the number of meshpoints, we can decrease the
viscosity by a factor of about $2^5=32$.
This shows that hyperviscosity can allow
a dramatic increase of the Reynolds number based on the scale of the box.

Higher order hyperviscosities ($n=3$ and larger) have been studied
previously \cite{BO95,CLV02}, but for us
$n=3$ is a practical limit, because we have restricted
the maximum stencil length of all derivative schemes to three (in each
direction), which is required for sixth order finite difference schemes
for our first and second derivatives \cite{B03}.

In the following we consider the convergence of the energy spectrum
for our hyperviscous simulations and compare with direct simulations.
We discuss then the Reynolds number dependence of the normalized mean
dissipation rate, and present finally the scaling behavior of the
structure functions.
Our basic conclusion is that in hyperviscous and direct simulations,
as well as in wind tunnel experiments, the inertial range scaling is
virtually identical and the width of the bottleneck is similar.

\subsection{Energy spectra}

Here and below we have calculated the energy dissipation rate from the
energy spectrum via
\EQ
\epsilon=2\nu_n\int k_{\rm eff}^{2n}E(k)\,\dd k.
\label{epsEk}
\EN
Here we have taken into account that in the code we employ a
finite difference scheme which has always a discretization error,
so we have to use the effective wavenumber in the expression above.
The effective wavenumber is usually less than the actual one;
see figure~9.1 of Ref.~\cite{B03}.
For example, for the sixth order finite difference scheme, an analytic
expression for $k_{\rm eff}^2$ was given in Ref.~\cite{CHB01}, while in
the present case we have
\EQ
\kappa_{\rm eff}^6
=20-30\cos\kappa+12\cos2\kappa-2\cos3\kappa,
\EN
where $\kappa=k\delta x$ is the wavenumber scaled by the mesh spacing
$\delta x$.
Using the effective wavenumber becomes particularly important in
the hyperviscous case in order not to overestimate the contribution
to $\epsilon$ in \Eq{epsEk}.

\begin{figure}[t!]
  \centering
  \includegraphics[width=.5\textwidth]{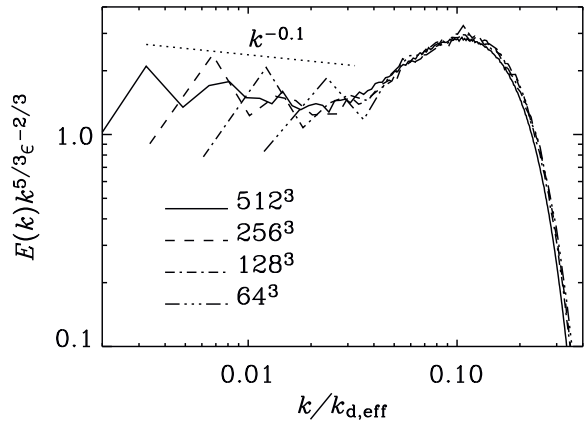}
  \caption{Time-averaged energy spectra
compensated by $k^{-5/3}\epsilon^{-2/3}$.
The curves correspond to four different resolutions.
All runs are with hyperviscosity.}
 \label{hydro}
\end{figure}

\begin{figure}[t!]
  \centering
  \includegraphics[width=.5\textwidth]{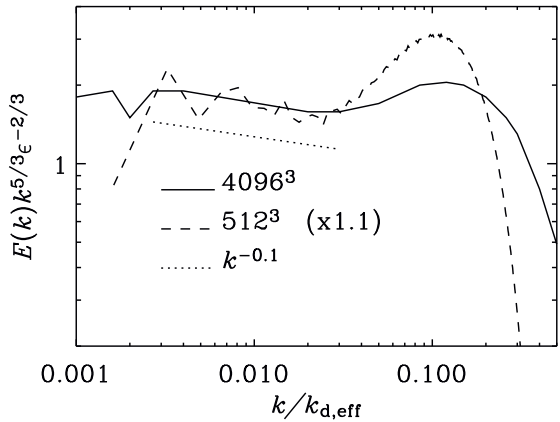}
  \caption{
Time-averaged energy spectra, compensated by $k^{5/3}\epsilon^{-2/3}$,
for the direct simulation with $4096^3$ meshpoints at $\Rey_{\lambda}=1201$
(solid line) from figure~5 of Ref.~\cite{Kan03} and our hyperviscous
simulation with $512^3$ meshpoints (dashed line).
Note that the bottleneck has a higher amplitude in the
hyperviscous case, but the inertial range has the same
slope as for the simulation with $4096^3$ meshpoints. 
Our hyperviscous energy spectrum is scaled by a factor 1.1 in order to 
make it fall on top of the $4096^3$ result, i.e.\ our Kolmogorov constant
is 1.1 times smaller than for the $4096^3$ simulation.
  }
 \label{kaneda_comp_hyper}
\end{figure}

\begin{figure}[t!]
  \centering
  \includegraphics[width=.5\textwidth]{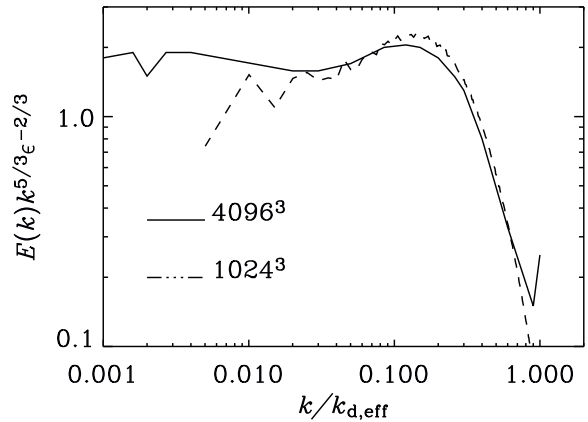}
  \caption{Our simulation with $1024^3$ meshpoints 
and normal viscosity show a bottleneck
very similar to the bottlenecks in Ref~\cite{Kan03}, but due to lack of
resolution we do not see any inertial range.
  }
 \label{kaneda_comp_normal}
\end{figure}

The dissipation wavenumber, $k_{\rm d}$, is calculated from the
relations $\epsilon=k_{\rm d} u_{k_{\rm d}}^3$ and
$\epsilon=\nu_n k_{\rm d}^{2n} u_{k_{\rm d}}^2$.
This leads to
\EQ
k_{\rm d}^{6n-2}=\epsilon/\nu_n^3\quad
\left(=k_{\rm d}^{16}\;\;\mbox{for}\;n=3\right).
\EN
Again, for $n=1$ one recovers the usual relation
$k_{\rm d}=(\epsilon/\nu^3)^{1/4}$.
For larger values of $n$ we find that, in order to make the location of the
inertial range in direct and hyperviscous simulations agree, we have to use
an effective wavenumber $k_{\rm d,eff}$ that is larger than $k_{\rm d}$
by a factor that is around 4 in our case,
i.e.\ $k_{\rm d,eff} \approx 4 k_{\rm d}$.

In \Fig{hydro} we show the convergence of the energy spectra of
hyperviscous runs for increasing resolution up to $512^3$ meshpoints.
All spectra are compensated by a $k^{5/3}\epsilon^{-2/3}$ factor and
the abscissa is normalized to the effective dissipation wavenumber 
$k_{\rm d,eff}$.
All runs agree in the shape of the bottleneck and the subsequent
dissipation subrange, but the length of the inertial range varies
from non-existent to about one order of magnitude.

We now compare our $512^3$ meshpoints hyperviscous run with the
direct simulations of Kaneda et al.\ \cite{Kan03} on the Earth Simulator
using $4096^3$ meshpoints; see \Fig{kaneda_comp_hyper}.
We see that in both cases the bottleneck sets in at 
$k/k_{\rm d,eff}\approx0.03$ and spans approximately one decade,
but the dissipation subrange is longer in the direct simulations. 
The height of the bottleneck increases with increasing order of
the hyperviscosity \cite{BO95},
which is not surprising given that the
steepness of the dissipative subrange is the reason for the
bottleneck effect in the first place \cite{bottleneck}.
In agreement with Kaneda et al.\ \cite{Kan03}, we find that
the slope of the energy spectrum in the inertial range
is consistent with the $k^{-1.77}$ law found in the direct simulation.
The Kolmogorov constant is however slightly smaller (about $\times1.1$)
in our hyperviscous case.

We should emphasize that, although we solve the compressible equations
using finite differences, our direct simulations agree favorably with
those using spectral methods solving the incompressible equations.
This is shown in \Fig{kaneda_comp_normal} where we compare simulations
using $1024^3$ meshpoints and normal viscosity with those of Ref~\cite{Kan03}.
These data have previously been discussed in Refs.~\cite{DHYB03,HBD04}
in connection with the bottleneck effect in hydrodynamics and
hydromagnetic turbulence.

\begin{figure}[t!]
  \centering
  \includegraphics[width=.5\textwidth]{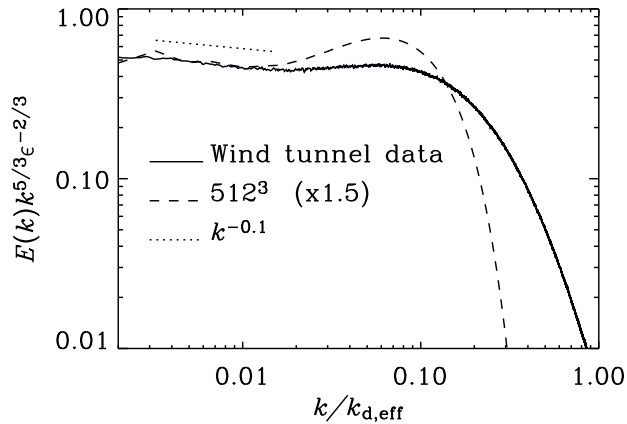}
  \caption{
One-dimensional time-averaged energy spectra of our largest run with 
hyperviscosity compared with wind tunnel data with 
$\Rey_\lambda=730$ \cite{PKJ03}. We have multiplied our energy spectra
by 1.5 in order to make it fall on top of the wind tunnel data.}
 \label{wind_comp_hyper}
\end{figure}

We now compare with the data from a wind tunnel experiment.
Ideally we would like to translate the one-dimensional wind tunnel
data into three-dimensional data \cite{DHYB03}, but this involves
differentiation which amplifies the noise in the data.
Therefore we now compare one-dimensional energy spectra of 
our largest hyperviscous simulation with the energy spectrum from a 
wind tunnel experiment; see \Fig{wind_comp_hyper}.
We see that in our simulation the bottleneck has larger amplitude
than in the wind tunnel experiment, but the (negative) slope of the
inertial range spectrum is comparable in the two cases, i.e.\ $1.77$.
The Kolmogorov constant on the other hand is smaller
by a factor of 1.5
in the hyperviscous case compared to the wind tunnel experiment.

We feel that the value of a slope of 1.77 should be taken with caution,
because it departs rather markedly from the value 1.70 expected from the
She-Leveque relation \cite{SL94}.
Given that the inertial range is still relatively short, a slope of 1.70
can certainly not be excluded.

It is customary to quote the Reynolds number based on the Taylor microscale
\cite{S98},
\EQ
\lambda=\sqrt{5}\,u_{\rm rms}/\omega_{\rm rms}.
\EN
Furthermore, $u_{\rm rms}$ and $\omega_{\rm rms}$ are the rms velocity and
vorticity, respectively.
One usually takes the one-dimensional rms velocity
for defining the Reynolds number,
\EQ
\label{Re_lambda}
\Rey_\lambda=u_{\rm1D}\lambda/\nu,
\EN
where $u_{\rm1D}^2={1\over3}u_{\rm rms}^2$.
The wind tunnel experiments have $\Rey_{\lambda}=730$.

In the hyperviscous case the straightforward
definition of the Taylor microscale Reynolds number would be
$\Rey_\lambda=u_{\rm1D}\lambda^5/\nu_3$, but this would lead to rather
large values ($\sim10^6$) which would not be meaningful in this context.
Instead we define an effective viscosity from the actual mean dissipation
rate and the modulus of the ordinary rate of strain matrix,
\EQ
\label{nu_eff}
\nu_{\rm eff}=\bra{\epsilon}/\bra{2\SSSS^2},
\EN
which is then used to estimate the value of $\nu$ in \Eq{Re_lambda}. 
In this way we find $\Rey_{\lambda}=340$ for our largest simulation.
Comparing with the high resolution direct simulations (\Fig{kaneda_comp_hyper})
and with wind tunnel data (\Fig{wind_comp_hyper})
we see that  $\Rey_{\lambda}=340$ probably is an underestimate
for our hyperviscous simulations.

Alternatively one can define $\Rey_\lambda$ as a measure of the width of the
inertial range.
Using relations that are valid in the standard case with $n=1$, we have
$k_{\rm d,eff}/k_{\rm f}\sim\Rey^{3/4}$ and
$\Rey_\lambda\sim\Rey^{1/2}$, which yields
\EQ
\label{Re_l}
\Rey_{\lambda}\approx\Rey_{\lambda0}
\left(\frac{k_{\rm d,eff}}{k_{\rm f}}\right)^{2/3},
\EN
where we have introduced $\Rey_{\lambda0}$ as a calibration parameter,
and $k_{\rm f}$ is the forcing wavenumber or, more generally, the wavenumber
of the energy carrying scale.
If we set $\Rey_{\lambda0}\approx7.5$, we can reproduce the result
$\Rey_{\lambda}=340$ for our largest run.
On the other hand, if we choose to calibrate $\Rey_{\lambda0}$ such that
our run with $512^3$ meshpoints and the wind tunnel experiments
have the same $\Rey_{\lambda}=730$ (see; \Fig{wind_comp_hyper}) then
we find $\Rey_{\lambda0}=16$, which is perhaps a more reasonable estimate.

\subsection{Energy dissipation rate}

\begin{figure}[t!]
  \centering
  \includegraphics[width=.5\textwidth]{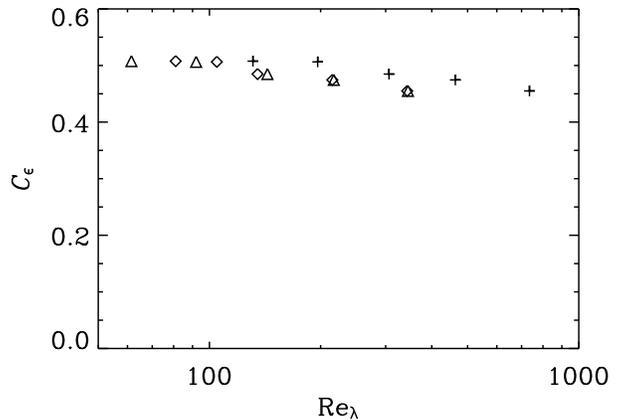}
  \caption{Plot of $C_\epsilon$ as a function of $\Rey_{\lambda}$ 
    for runs with third order hyperviscosity ($n=3$).
Triangles and plus signs represent
Reynolds numbers calculated based on \Eq{Re_l} with
$\Rey_{\lambda 0}=7.5$ and $16$, respectively,
while for the plus signs
\Eq{Re_lambda} together with \Eq{nu_eff} have been used.
  }
 \label{Ceps}
\end{figure}

According to the Kolmogorov phenomenology, the spectral energy flux should
be independent of $k$ in the inertial range and equal to both the rate
of energy input at large scales and the rate of energy dissipation at
small scales.
The constant of proportionality is of fundamental interest in turbulence
research and one wants to know whether this value
is independent of Reynolds number \cite{S98,Kan03}.
It is customary to define this coefficient as
\EQ
C_{\epsilon}=\bra{\epsilon}L/u_{\rm 1D}^3,
\EN
where $\bra{\epsilon}$ is the mean energy dissipation rate and $L$ the
integral scale, which is usually defined as $L=(3\pi/4)k_{\rm I}^{-1}$,
where
\EQ
k_{\rm I}^{-1}=\left.\int k^{-1}E(k)\,\dd k
\right/\int E(k)\,\dd k,
\EN
is the spectrally weighted average of $k^{-1}$.

The resulting normalized mean energy dissipation rate, calculated in
this way, is shown in \Fig{Ceps} as a function of Reynolds number.
In the figure diamonds correspond to using \Eq{Re_lambda} and \Eq{nu_eff} 
to find the
Reynolds number, while triangles and plus signs correspond to using 
\Eq{Re_l} with $\Rey_{\lambda 0}=7.5$ and $16$, respectively.
\FFig{Ceps} shows that
our results are in good agreement with both numerical \cite{Kan03,PYHBK04}
and experimental \cite{S98} data.

\subsection{Structure functions}

\begin{figure}[t!]
  \centering
  \includegraphics[width=.5\textwidth]{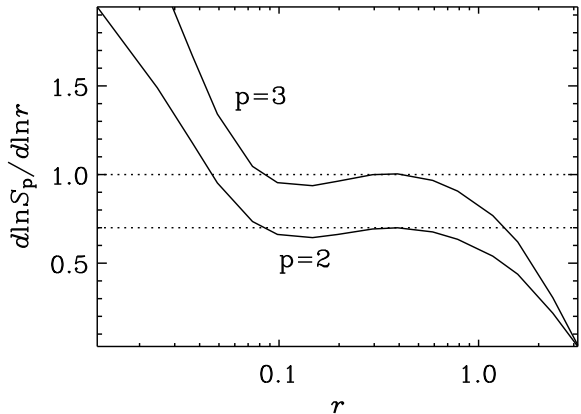}
  \caption{
Time-averaged
total structure functions, $[S_p^{(l)}+2S_p^{(t)}]/3$, for $p=2$ and $p=3$.
The two dotted horizontal lines go through 0.7 and 1.0, confirming the
expected scaling from the She-Leveque relationship.
  }
 \label{structure}
\end{figure}

\begin{figure}[t!]
  \centering
  \includegraphics[width=.5\textwidth]{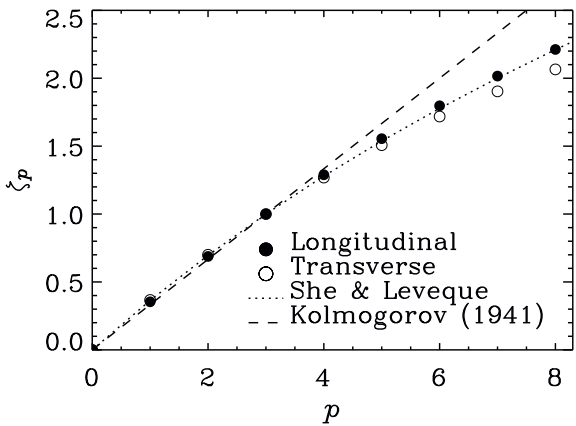}
  \caption{
Structure function scaling exponents found using the concept of 
extended self similarity. We see that the longitudinal 
scaling exponents follow the She \& Leveque scaling very well, while
the transversal scaling exponents are somewhat more intermittent.
  }
 \label{slope_struct}
\end{figure}

The spectral information can be supplemented by similar
scaling information in real space using structure functions.
We define the longitudinal and transversal structure functions
\EQ
S_p^{\rm(l)}(r)
=\left\langle\left\{\rr\cdot\left[\uu(\xx+\rrr)-\uu(\xx)
\right] \right\}^p \right\rangle,
\EN
\EQ
S_p^{\rm(t)}(r)
=\left\langle\left\{\nnn\cdot\left[\uu(\xx+\rrr)-\uu(\xx)
\right]\right\}^p \right\rangle,
\EN
respectively.
Here, $\rr$ is the unit vector of $\rrr$ and $\nnn$ is normal to $\rrr$,
so $\nnn \cdot \rr \equiv 0$.
The structure function of the three-dimensional velocity field,
\EQ
S_p(r)
=\langle |\uu(\xx+\rrr)-\uu(\xx)|^p \rangle,
\EN
can then be written as
\EQ
S_p(r)={\textstyle\frac{1}{3}}[S_p^{\rm(l)}(r)+2S_p^{\rm(t)}(r)].
\EN
We define the $p$th order structure function scaling exponent, 
$\zeta_p$ via the scaling relation
\EQ
S_p(r) \propto r^{\zeta_p}.
\EN
In \Fig{structure} we plot the derivative of the double-logarithmic slope
of the structure functions, $\dd\ln S_p/\dd\ln r$.
Inertial range scaling is indicated by a plateau in this graph.
We find from the lower curve of \Fig{structure} that $\zeta_2 \approx 0.7$.
More importantly, in the upper curve of \Fig{structure}, we show that 
$S_3(r)$ is consistent with linear
scaling, i.e. $\zeta_3=1$.
Knowing this we can use the
extended self similarity \cite{Benzi93} 
to find the other structure function
scaling exponents. The results are shown in \Fig{slope_struct} where we see
that the longitudinal structure function exponents
follow the She \& Leveque \cite{SL94}
scaling very well, while the transversal structure function is slightly 
more intermittent (i.e.\ the graph of $\zeta_p$ versus $p$ is more strongly bent).
In particular we note that the extended self similarity gives $\zeta_2 = 0.696$
for the longitudinal component.

\subsection{Decaying turbulence}

\begin{figure}[t!]
  \centering
  \includegraphics[width=.5\textwidth]{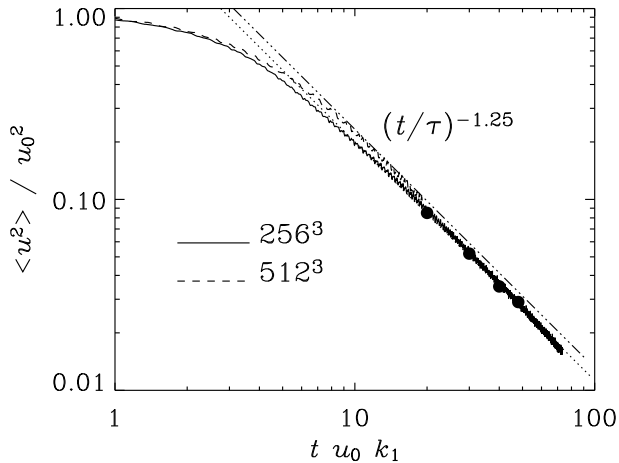}
  \caption{
Mean squared velocity as a function of time for a simulation
where the forcing has been stopped at $t=t_0$.
In the $256^3$ simulation we have $\Rey_{\rm grid}=0.20$ (based on the
initial value $u_0$) and
$\tau u_0 k_1=2.8$ while in the $512^3$ simulation we have
$\Rey_{\rm grid}=0.24$ and $\tau u_0 k_1=3.1$.
The solid circles correspond to the experimental results of
Kang et al.\ \cite{KCM03}.
  }
 \label{pdecay_comp}
\end{figure}

Decaying isotropic turbulence is often considered an important benchmark
of turbulence theories, and comparisons with wind tunnel experiments
and large eddy simulations are available \cite{KCM03}.
We have carried out simulations of decaying turbulence by stopping the
driving at a time that will now be redefined to $t=0$.
In \Fig{pdecay_comp} we show that asymptotically
\EQ
\bra{\uu^2}/u_0^2=(t/\tau)^n,
\EN
where $u_0$ is the initial ($t=0$) rms velocity and $\tau$ is a
constant obtained from the intersection of the decay law extrapolated
to $\bra{\uu^2}=u_0^2$.
It turns out that our law is consistent with $n=1.25$, which is in
agreement with recent wind tunnel data and large eddy simulations
\cite{KCM03}.

\begin{figure}[t!]
  \centering
  \includegraphics[width=.5\textwidth]{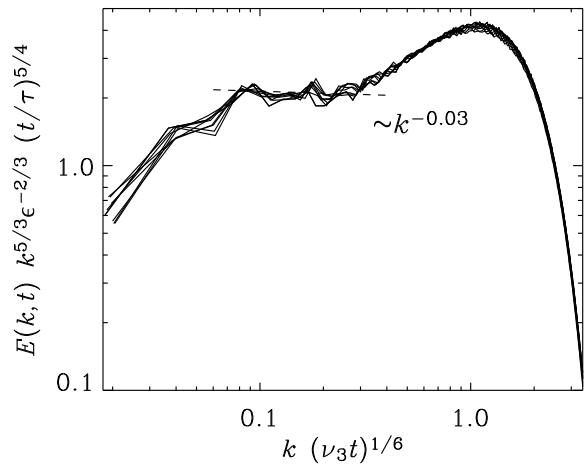}
  \caption{
Energy spectra for a decaying run.
The abscissa is compensated by $(\nu_3t)^{1/6}$ to make it dimensionless
and to account for the slow decrease of the dissipation wavenumber.
The ordinate is compensated by $k^{5/3}$ to show the location of the
inertial range and by $t^{5/4}$ to compensate for the decay.
  }
 \label{pspec_comp}
\end{figure}

The bottleneck effect is roughly unchanged; see \Fig{pspec_comp}.
Its width is still about one order of magnitude in wavenumber,
which is comparable to the experimental results \cite{KCM03}.
The height of the bottleneck is much less in the experimental
data, because they show only a one-dimensional spectrum which
gives a much weaker hump than the three-dimensional spectra
\cite{DHYB03}.
In any case, we know already that the height of the bottleneck
is artificially enhanced by the use of hyperviscosity.

In the present decay simulations one also sees the beginning of a
subinertial range, leaving only a very short inertial range around
$0.1\la k(\nu_3t)^{1/6}\la 0.3$.
The slope is compatible with the She-Leveque slope of 1.70, which
corresponds to a residual slope of $k^{-0.03}$ after compensating with
$k^{5/3}$.
Thus, there is no longer evidence for the more extreme correction of
$-0.1$ suggested by the forced turbulence simulations \cite{Kan03}.

\section{Conclusion}

The present investigations have shown that turbulence simulations with
hyperviscosity are able to reproduce virtually the same inertial range
scalings as simulations with ordinary viscosity.
Specifically, the structure function exponents show scaling behavior
that is consistent with the She-Leveque \cite{SL94} model.
However, the transversal structure functions show a slightly higher
degree of intermittency than the longitudinal ones.
This, in turn, is quite consistent with a number of turbulence
simulations by other groups \cite{gotoh02,PPW02}.
A possible explanation for the difference between longitudinal and transversal
structure functions has been offered by Siefert \& Peinke \cite{SP03},
who find different cascade times for longitudinal and transversal spectra.
The spectra show inertial range scaling similar to that found both in
wind tunnel experiments \cite{PKJ03} and in very high resolution direct
simulations \cite{Kan03}.
In all three cases (hyperviscous and direct simulations as well as wind
tunnel experiments) the inertial range spectrum is found to be compatible
with the $k^{-1.77}$ behavior found by Kaneda et al.\ \cite{Kan03}.
As discussed above, this result is not compatible with the results from
the structure function scalings and the She-Leveque relation.
However, we believe that the presently resolved inertial range is still
too short to distinguish conclusively between 1.77 and the She-Leveque
value of 1.70.
Also, the simulation data of decaying turbulence suggest a weaker
correction of 0.03, giving a slope of 1.70 that is compatible with the
She-Leveque scaling.

Another important result is that the width of the bottleneck seems to
be independent of the use of hyperviscosity, and that only its height
increases with the order of the hyperviscosity.
This result is also confirmed in the case of decaying turbulence.
Finally, we note that the normalized dissipation rate is independent of
the Reynolds number, and that the asymptotic value of 
$C_{\epsilon}\approx 0.5$ is
in agreement with both experimental and numerical results \cite{PKW02,Kan03}.

One should of course always be concerned about the possible side effects
of using hyperviscosity.
One worry is that hyperviscosity may actually affect almost all
of the inertial subrange \cite{BSC98,BM00}.
The current simulations confirm that the bottleneck requires at least
an order of magnitude in $k$-space, and so does the dissipative subrange,
leaving almost no inertial range at all--even in a simulation with $1024^3$
meshpoints.
Thus, using hyperviscosity appears to be a reasonable procedure for gaining
information about the inertial range at moderate cost, even though one
should still use a reasonably high resolution to isolate true inertial range
features from those in the bottleneck subrange.
On the other hand, hyperviscosity is not a universally valid approximation.
An example is in magnetohydrodynamics when magnetic helicity is finite
and a large scale magnetic field builds up in a closed or fully periodic
box \cite{BS02}.
As long as it is possible to understand the
origin of peculiar features arising from hyperviscosity or
hyper-resistivity (as is the case in helical hydromagnetic turbulence)
there may well be circumstances where turbulence with hyperviscosity can
provide a useful model for certain studies.
One should bear in mind, however, that the height of the bottleneck depends 
on the order of the hyperviscosity. 
For example, if we choose $n > 3$ in \Eq{22prelim}, the 
height of the bottleneck will be even more exaggerated \cite{BO95}.

\acknowledgments
We acknowledge the very valuable help and
discussions with B. R. Pearson.
We are also indebted to an anonymous referee for making
useful suggestions regarding the study of decaying turbulence.
We thank the Danish Center
for Scientific Computing for granting time on the Horseshoe cluster,
and the Norwegian High Performance Computing Consortium (NOTUR)
for granting time on the parallel computers in 
Trondheim (Gridur/Embla) and Bergen (Fire).


\end{document}